# *A multiple-relaxation-time lattice Boltzmann model for simulating incompressible axisymmetric thermal flows in porous media*

Qing Liu[a], Ya-Ling He[a], Qing Li[b]

[a]*Key Laboratory of Thermo-Fluid Science and Engineering of Ministry of Education, School of Energy and Power Engineering, Xi'an Jiaotong University, Xi'an, Shaanxi, 710049, China*

[b]*School of Energy Science and Engineering, Central South University, Changsha 410083, China*

**Abstract**

In this paper, a multiple-relaxation-time (MRT) lattice Boltzmann (LB) model is developed for simulating incompressible axisymmetric thermal flows in porous media at the representative elementary volume (REV) scale. In the model, a D2Q9 MRT-LB equation is proposed to solve the flow field in addition to the D2Q5 LB equation for the temperature field. The source terms of the model are simple and contain no velocity and temperature gradient terms. The generalized axisymmetric Navier-Stokes equations for axisymmetric flows in porous media are correctly recovered from the MRT-LB model through the Chapman-Enskog analysis in the moment space. The present model is validated by numerical simulations of several typical axisymmetric thermal problems in porous media. The numerical results agree well with the data reported in the literature, demonstrating the effectiveness and accuracy of the present MRT-LB model for simulating axisymmetric thermal flows in porous media.

# 1. Introduction

The analysis of axisymmetric convective transport processes in porous media has attracted considerable attention due to its importance in many areas such as geothermal energy systems, contaminant transport in groundwater, underground treatment of nuclear waste materials, fibrous insulation, crude oil extraction, chemical catalytic reactors, electronic device cooling, and porous heat exchangers, to name only a few applications. Comprehensive literature surveys concerning this topic have been given in the books by Nield and Bejan [1], and Vafai [2]. In the past several decades, axisymmetric fluid flow and heat transfer problems in porous media at the representative elementary volume (REV) scale have been numerically investigated by many researchers using some traditional numerical techniques, such as the finite difference method [3-5], finite volume method [6, 7], and finite element method [8].

The lattice Boltzmann (LB) method, as a mesoscopic numerical technique originates from the lattice-gas automata (LGA) method [9], has become an effective computational method for simulating complex fluid flows and modeling complex physics in fluids owing to its kinetic background [10-18]. In the past decade, the LB method for axisymmetric fluid flow and heat transfer problems has also attracted much attention. Generally speaking, three-dimensional axisymmetric fluid flow and heat transfer problems can be reduced to quasi-two-dimensional ones in cylindrical coordinate system, which greatly enhances the computational efficiency. To make full use of this feature, Halliday et al. [19] proposed the first axisymmetric LB model for incompressible axisymmetric flows by introducing the velocity and spatial dependent source terms into the LB equation to account for the additional terms arising from the cylindrical coordinate system. Since then, many axisymmetric LB models have been developed to study axisymmetric isothermal flows [20-29] and thermal flows [30-33]. The above mentioned axisymmetric LB models are limited to axisymmetric fluid flow and heat transfer problems in the absence of porous media. Recently, Rong et al. [34] proposed an axisymmetric LB model for axisymmetric convection heat transfer in porous media at the REV scale. The axisymmetric LB model [34] employs the Bhatnagar-Gross-Krook (BGK) collision operator and can be used to simulate axisymmetric flows in a medium with constant or variable porosity. However, to the best of the authors' knowledge, no reports have been found in the literature about the applications of the multiple-relaxation-time (MRT) LB method to axisymmetric flows in porous media at the REV scale. In the LB community, it has been widely accepted that the MRT-LB model [35, 36] has better numerical stability and accuracy than the BGK-LB model [37]. As reported in Refs. [25, 26], the numerical stability of the axisymmetric MRT-LB model is much better than the axisymmetric BGK-LB model at low viscosities. This motivates the present work, where the main purpose is to develop an axisymmetric MRT-LB model for simulating incompressible axisymmetric thermal flows in porous media based on some previous studies [24, 25, 32, 38]. In the model, a D2Q9 MRT-LB equation is proposed to solve the flow field of the axisymmetric thermal flows in porous media in the framework of the standard MRT-LB method. The porosity is included into the equilibrium moments, and the linear and nonlinear drag forces of the porous matrix are incorporated into the model by adding a source term into the MRT-LB equation in the moment space.

The rest of this paper is organized as follows. In Section 2, the macroscopic governing equations for axisymmetric thermal flows in porous media are briefly introduced. In Section 3, the MRT-LB model for axisymmetric thermal flows in porous media is presented in detail. In Section 4, numerical simulations of several typical axisymmetric thermal flows in porous media are carried out to validate the effectiveness and accuracy of the present MRT-LB model. Finally, Section 5 concludes the paper.

**2. Macroscopic governing equations**

The fluid flow is assumed to be laminar, incompressible, and Newtonian without viscous heat dissipation and pressure work. The Brinkman-Forchheimer-extended Darcy model (also called the generalized model) [39] and the local thermal equilibrium model [1] are employed to describe the momentum and energy transfer in porous media, respectively. In cylindrical coordinate system, the macroscopic governing equations for incompressible axisymmetric thermal flows in homogeneous and isotropic porous media at the REV scale are given as follows [39]:

$$\frac{1}{r}\frac{\partial(ru_r)}{\partial r}+\frac{\partial u_z}{\partial z}=0, \tag{1}$$

$$\frac{\partial u_r}{\partial t}+\frac{1}{r}\frac{\partial}{\partial r}\left(\frac{ru_r u_r}{\phi}\right)+\frac{\partial}{\partial z}\left(\frac{u_z u_r}{\phi}\right)=-\frac{1}{\rho}\frac{\partial(\phi p)}{\partial r}+\upsilon_e\left(\nabla^2 u_r-\frac{u_r}{r^2}\right)+F_r, \tag{2}$$

$$\frac{\partial u_z}{\partial t}+\frac{1}{r}\frac{\partial}{\partial r}\left(\frac{ru_z u_r}{\phi}\right)+\frac{\partial}{\partial z}\left(\frac{u_z u_z}{\phi}\right)=-\frac{1}{\rho}\frac{\partial(\phi p)}{\partial z}+\upsilon_e\nabla^2 u_z+F_z, \tag{3}$$

$$\frac{\partial T}{\partial t}+u_r\frac{\partial T}{\partial r}+u_z\frac{\partial T}{\partial z}=\alpha_e\nabla^2 T, \tag{4}$$

where

$$\nabla^2=\frac{1}{r}\frac{\partial}{\partial r}\left(r\frac{\partial}{\partial r}\right)+\frac{\partial^2}{\partial z^2}, \tag{5}$$

$u_r$ and $u_z$ are the components of the velocity **u** in the *r*- (radial) and *z*- (axial) directions, respectively, $\rho$ is the fluid density, $p$ is the pressure, $T$ is the temperature, $\phi$ is the porosity of the porous media, $\upsilon_e$ and $\alpha_e$ are the effective kinetic viscosity and thermal diffusivity of the porous media, respectively. The total body force $\mathbf{F}=(F_r, F_z)$, which is induced by the solid matrix and other external forces, can be expressed as [40, 41]

$$\mathbf{F}=-\frac{\phi\upsilon}{K}\mathbf{u}-\frac{\phi F_\phi}{\sqrt{K}}|\mathbf{u}|\mathbf{u}+\phi\mathbf{G}, \tag{6}$$

where $\upsilon$ is the kinetic viscosity of the fluid, and $|\mathbf{u}|=\sqrt{u_r^2+u_z^2}$. Based on the Boussinesq approximation, **G** is given by

$$\mathbf{G}=g\beta(T-T_0)\mathbf{j}+\mathbf{a}, \tag{7}$$

where $g$ is the gravitational acceleration, $\beta$ is the thermal expansion coefficient, $T_0$ is the reference temperature, **j** is the unit vector in the *z*-direction, and **a** is the acceleration induced by other external forces. According to Ergun's experimental relation [42], the geometric function $F_\phi$ and the permeability $K$ of the porous media can be estimated in terms of the porosity $\phi$ and the particle diameter $d_p$ as [43]

$$F_\phi = \frac{1.75}{\sqrt{150\phi^3}}, \quad K = \frac{\phi^3 d_p^2}{150(1-\phi)^2}. \tag{8}$$

The flow governed by Eqs. (1)-(4) is characterized by $\phi$ and several dimensionless parameters: the Rayleigh number $Ra$, the Darcy number $Da$, the Prandtl number $Pr$, the Reynolds number $Re$, the viscosity ratio $J$, and the thermal diffusivity ratio $\lambda$, which are defined as follows

$$Ra = \frac{g\beta\Delta T L^3}{\upsilon\alpha}, \quad Da = \frac{K}{L^2}, \quad Pr = \frac{\upsilon}{\alpha}, \quad Re = \frac{LU}{\upsilon}, \quad J = \frac{\upsilon_e}{\upsilon}, \quad \lambda = \frac{\alpha_e}{\alpha}, \tag{9}$$

where $\Delta T$ is the temperature difference (characteristic temperature), $L$ is the characteristic length, $U$ is the characteristic velocity, and $\alpha$ is the thermal diffusivity of the fluid.

The first term and the second term on the right-hand-side of Eq. (6) are the Darcy (linear) and Forchheimer (nonlinear) drag forces due to the solid matrix, respectively. The nonlinear drag force term can be neglected for a small Darcy number $Da$ or Reynolds number $Re$. However, for large Darcy number $Da$ or Reynolds number $Re$, the influence of the nonlinear drag force must be considered [41]. As $\phi \to 1$ and $Da \to \infty$, the generalized axisymmetric Navier-Stokes equations (1)-(3) reduce to the axisymmetric Navier-Stokes equations for free fluid flows (without porous media). In addition, when $F_\phi = 0$ (the nonlinear drag force is neglected), Eqs. (2) and (3) reduce to the Brinkman-extended Darcy equation.

## 3. MRT-LB model for axisymmetric thermal flows in porous media

The macroscopic governing equations (1)-(4) for incompressible axisymmetric thermal flows in porous media can be rewritten as follows:

$$\partial_j u_j = -u_r/r, \tag{10}$$

$$\rho_0 \left[ \partial_t u_j + \partial_k \left( \frac{u_j u_k}{\phi} \right) \right] = -\partial_j (\phi p) + \mu \partial_k \left( \partial_k u_j + \partial_j u_k \right) + \underline{\frac{\mu}{r} \left( \partial_r u_j + \partial_j u_r \right) - \frac{\rho u_j u_r}{r \phi} - \frac{2\mu u_j \delta_{jr}}{r^2}} + \rho F_j , \quad (11)$$

$$\partial_t T + \partial_j (u_j T) = \partial_j (\alpha_e \partial_j T) + \underline{\alpha_e \frac{1}{r} \partial_r T - \frac{u_r T}{r}} , \quad (12)$$

where $j$, $k$ indicate the $r$ or $z$ component, $\mu = \rho \upsilon_e$ is the effective dynamic viscosity, and $\delta_{jr}$ is the Kronecker $\delta$ function with two indices. The underline terms in Eqs. (11) and (12) arise from the cylindrical coordinate system. In the present work, the methods reported in Refs. [24, 32] are adopted to recover these terms in the macroscopic equations. In what follows, the axisymmetric MRT-LB model for axisymmetric thermal flows in porous media is presented in detail.

*3.1 D2Q9 MRT-LB equation for the flow field*

It has been widely accepted that the MRT collision model can enhance the numerical stability and accuracy of the LB scheme as compared with the BGK collision model. The dimensionless relaxation times of the conserved (hydrodynamic) and nonconserved (kinetic) moments can be adjusted independently when using the MRT collision model [25, 26, 35, 36, 44-46]. According to Refs. [26, 44], we introduce the following D2Q9 MRT-LB equation to solve the flow field:

$$\tilde{f}_i (\mathbf{x} + \mathbf{e}_i \delta_t, t + \delta_t) - \tilde{f}_i (\mathbf{x}, t) = -\tilde{\Lambda}_{i\beta} \left[ \tilde{f}_\beta - f_\beta^{(eq)} \right]\bigg|_{(\mathbf{x},t)} + \frac{\delta_t}{2} \left[ \tilde{S}_i (\mathbf{x}, t) + \tilde{S}_i (\mathbf{x} + \mathbf{e}_i \delta_t, t + \delta_t) \right], \quad (13)$$

where $\tilde{f}_i (\mathbf{x}, t)$ is the volume-averaged density distribution function with velocity $\mathbf{e}_i = (e_{ir}, e_{iz})$ at position $\mathbf{x} = (r, z)$ and time $t$, $f_i^{(eq)} (\mathbf{x}, t)$ is the equilibrium distribution function, $\tilde{S}_i$ is the source term accounting for the total body force and the additional terms arising from the cylindrical coordinate system, and $\tilde{\Lambda} = \mathbf{M}^{-1} \Lambda \mathbf{M}$ is the collision matrix, in which $\mathbf{M}$ is a $9 \times 9$ orthogonal transformation matrix and $\Lambda = \mathbf{M} \tilde{\Lambda} \mathbf{M}^{-1} = \text{diag}(s_0, s_1, \cdots, s_8)$ is a $9 \times 9$ diagonal relaxation matrix ($\{s_i \mid 0 < s_i < 2\}$ are dimensionless relaxation rates).

The nine discrete velocities $\{\mathbf{e}_i \mid i = 0, 1, \ldots, 8\}$ of the D2Q9 model are given by [37]

$$\mathbf{e}_i = \begin{cases} (0,0), & i = 0 \\ \left(\cos\left[(i-1)\pi/2\right], \sin\left[(i-1)\pi/2\right]\right)c, & i = 1 \sim 4 \\ \left(\cos\left[(2i-1)\pi/4\right], \sin\left[(2i-1)\pi/4\right]\right)\sqrt{2}c, & i = 5 \sim 8 \end{cases} \quad (14)$$

where $c = \delta_r/\delta_t$ is the lattice speed with $\delta_r$ and $\delta_t$ representing the lattice spacing and time step, respectively. In the present LB model, $c$ is set to be 1 (i.e., $\delta_r = \delta_t$).

The implicitness of Eq. (13) can be eliminated by using a new distribution function $f_i = \tilde{f}_i - 0.5\delta_t \tilde{S}_i$, from which the following explicit D2Q9 MRT-LB equation can be obtained:

$$\mathbf{f}(\mathbf{x}+\mathbf{e}\delta_t, t+\delta_t) - \mathbf{f}(\mathbf{x},t) = -\mathbf{M}^{-1}\mathbf{\Lambda}\left[\mathbf{m}(\mathbf{x},t) - \mathbf{m}^{(eq)}(\mathbf{x},t)\right] + \delta_t \mathbf{M}^{-1}\left(\mathbf{I} - \frac{\mathbf{\Lambda}}{2}\right)\mathbf{S}, \quad (15)$$

where $\mathbf{I}$ is the unit matrix, and the boldface symbols, $\mathbf{f}$, $\mathbf{m}$, $\mathbf{m}^{(eq)}$, and $\mathbf{S}$, are 9-dimensional column vectors, e.g., $\mathbf{f}(\mathbf{x},t) = (f_0(\mathbf{x},t), f_1(\mathbf{x},t), \cdots, f_8(\mathbf{x},t))^T$, in which T is the transpose operator, and $\mathbf{S} = \mathbf{M}\tilde{\mathbf{S}} = \mathbf{M}(\tilde{S}_0, \tilde{S}_1, \ldots, \tilde{S}_8)^T$ is the source term in the moment space.

The transformation matrix $\mathbf{M}$ linearly transforms the discrete distribution functions $\mathbf{f} \in \mathbb{V} = \mathbb{R}^9$ (velocity space) to the velocity moments $\mathbf{m} \in \mathbb{M} = \mathbb{R}^9$ (moment space):

$$\mathbf{m} = \mathbf{M}\mathbf{f}, \quad \mathbf{f} = \mathbf{M}^{-1}\mathbf{m}. \quad (16)$$

The nine velocity moments are given by

$$\mathbf{m} = (m_0, m_1, m_2, m_3, m_4, m_5, m_6, m_7, m_8)^T$$
$$= r(\rho, e, \varepsilon, j_r, q_r, j_z, q_z, p_{rr}, p_{rz})^T, \quad (17)$$

where $m_0 = \rho$ is the fluid density, $m_1 = e$ is related to the energy, $m_2 = \varepsilon$ is related to the energy square, $m_{3,5} = j_{r,z}$ are components of the momentum $\mathbf{J} = (j_r, j_z) = \rho\mathbf{u}$, $m_{4,6} = q_{r,z}$ are related to the energy flux, and $m_{7,8} = p_{rr,rz}$ are related to the diagonal and off-diagonal components of the stress tensor [36]. As reported in Ref. [25], the radial coordinate $r$ has been incorporated into the velocity moments (17) based on Guo et al.'s axisymmetric LB model [24]. With the ordering of the above specified velocity moments, the transformation matrix $\mathbf{M}$ is given by ($c=1$) [36]

$$\mathbf{M} = \begin{pmatrix} 1 & 1 & 1 & 1 & 1 & 1 & 1 & 1 & 1 \\ -4 & -1 & -1 & -1 & -1 & 2 & 2 & 2 & 2 \\ 4 & -2 & -2 & -2 & -2 & 1 & 1 & 1 & 1 \\ 0 & 1 & 0 & -1 & 0 & 1 & -1 & -1 & 1 \\ 0 & -2 & 0 & 2 & 0 & 1 & -1 & -1 & 1 \\ 0 & 0 & 1 & 0 & -1 & 1 & 1 & -1 & -1 \\ 0 & 0 & -2 & 0 & 2 & 1 & 1 & -1 & -1 \\ 0 & 1 & -1 & 1 & -1 & 0 & 0 & 0 & 0 \\ 0 & 0 & 0 & 0 & 0 & 1 & -1 & 1 & -1 \end{pmatrix}. \tag{18}$$

Among the nine velocity moments $\{m_i | i = 0, 1, \cdots, 8\}$, only the density $m_0 = \rho$ and momentum $m_{3,5} = j_{r,z}$ are conserved (hydrodynamic) quantities, the other velocity moments are nonconserved (kinetic) quantities. The equilibrium velocity moments $\mathbf{m}^{(eq)}$ for the velocity moments $\mathbf{m}$ are given by

$$\mathbf{m}^{(eq)} = r\left(\rho, e^{(eq)}, \varepsilon^{(eq)}, j_r, q_r^{(eq)}, j_z, q_z^{(eq)}, p_{rr}^{(eq)}, p_{rz}^{(eq)}\right)^{\mathrm{T}}. \tag{19}$$

where [38]

$$e^{(eq)} = -2\rho + \frac{3\rho(u_r^2 + u_z^2)}{\phi}, \quad \varepsilon^{(eq)} = \rho - \frac{3\rho(u_r^2 + u_z^2)}{\phi},$$

$$q_r^{(eq)} = -\rho u_r, \quad q_z^{(eq)} = -\rho u_z,$$

$$p_{rr}^{(eq)} = \frac{\rho(u_r^2 - u_z^2)}{\phi}, \quad p_{rz}^{(eq)} = \frac{\rho u_r u_z}{\phi}. \tag{20}$$

The evolution process of the MRT-LB equation (15) consists of two steps, i.e., the collision step and streaming step. Usually, the collision step is carried out in the moment space as

$$\mathbf{m}^+ = \mathbf{m} - \boldsymbol{\Lambda}\left[\mathbf{m} - \mathbf{m}^{(eq)}\right] + \delta_t \left(\mathbf{I} - \frac{\boldsymbol{\Lambda}}{2}\right)\mathbf{S}, \tag{21}$$

while the streaming step is still implemented in the velocity space

$$f_i(\mathbf{x} + \mathbf{e}_i \delta_t, t + \delta_t) = f_i^+(\mathbf{x}, t), \quad \mathbf{f}^+ = \mathbf{M}^{-1}\mathbf{m}^+. \tag{22}$$

The diagonal relaxation matrix $\boldsymbol{\Lambda}$ is given by

$$\boldsymbol{\Lambda} = \mathrm{diag}\left(1, s_e, s_\varepsilon, s_j, s_q, s_j, s_q, s_\upsilon, s_\upsilon\right). \tag{23}$$

The source term $\mathbf{S}$ in the moment space is given by [38]

$$\mathbf{S} = r\rho \begin{pmatrix} 0 \\ 6(u_r \tilde{F}_r + u_z \tilde{F}_z)/\phi \\ -6(u_r \tilde{F}_r + u_z \tilde{F}_z)/\phi \\ \tilde{F}_r \\ -\tilde{F}_r \\ \tilde{F}_z \\ -\tilde{F}_z \\ 2(u_r \tilde{F}_r - u_z \tilde{F}_z)/\phi \\ (u_r \tilde{F}_z + u_z \tilde{F}_r)/\phi \end{pmatrix}, \tag{24}$$

where $\tilde{\mathbf{F}} = (\tilde{F}_r, \tilde{F}_z)$ is given by

$$\tilde{F}_r = F_r + a_r, \quad \tilde{F}_z = F_z, \tag{25}$$

in which $a_r$ is defined as $a_r = c_s^2 [1 - 2(\tau_\upsilon - 0.5)\delta_t u_r/r]/r$ [24], $\tau_\upsilon$ is the dimensionless relaxation time for the flow field, and $c_s = c/\sqrt{3}$ is the sound speed of the D2Q9 model.

The fluid density $\rho$ and velocity $\mathbf{u} = (u_r, u_z)$ are defined as [24]

$$\rho = \frac{1}{r} \sum_{i=0}^{8} f_i, \tag{26}$$

$$\rho u_j = \frac{r}{r^2 + c_s^2 (\tau_\upsilon - 0.5) \delta_t^2 \delta_{jr}} \left[ \sum_{i=0}^{8} e_{ij} f_i + \frac{\delta_t}{2} r \rho F_j + \frac{\delta_t}{2} \rho c_s^2 \delta_{jr} \right]. \tag{27}$$

Note that the total body force $\mathbf{F} = (F_r, F_z)$ also contains the macroscopic velocity $\mathbf{u}$. According to Eqs. (6) and (27), the macroscopic velocity $\mathbf{u}$ can be calculated explicitly by

$$u_j = \frac{v_j}{l_{j0} + \sqrt{l_{j0}^2 + l_1 |\mathbf{v}|}}, \tag{28}$$

where $\mathbf{v} = (v_r, v_z)$, $l_{j0}$ and $l_1$ are given by

$$\rho v_j = \sum_{i=0}^{8} e_{ij} f_i + \frac{\delta_t}{2} r \phi \rho G_j + \frac{\delta_t}{2} \rho c_s^2 \delta_{jr}, \tag{29}$$

$$l_{j0} = \frac{1}{2} \left[ \frac{r^2 + c_s^2 (\tau_\upsilon - 0.5) \delta_t^2 \delta_{jr}}{r} + r\phi \frac{\delta_t}{2} \frac{\upsilon}{K} \right], \quad l_1 = r\phi \frac{\delta_t}{2} \frac{F_\phi}{\sqrt{K}}. \tag{30}$$

The pressure $p$ is defined as $p = \rho c_s^2 / \phi$, and the effective kinetic viscosity is defined as

$$\upsilon_e = c_s^2 \left( \tau_\upsilon - \frac{1}{2} \right) \delta_t \tag{31}$$

with $\tau_\upsilon = 1/s_\upsilon = 1/s_{7,8}$.

Through the Chapman-Enskog analysis in the moment space (see Appendix A for details), the generalized axisymmetric Navier-Stokes equations (10) and (11) can be recovered from the present MRT-LB model. It should be noted that, as $\phi \to 1$ and $Da \to \infty$, the present MRT-LB model reduces to the MRT-LB model [25] for axisymmetric flows in the absence of porous media. When the nine relaxation rates $\{s_i | 0 < s_i < 2\}$ are set to be a single value $1/\tau_\upsilon$, i.e., $\Lambda = (1/\tau_\upsilon)\mathbf{I}$, then the present MRT-LB model is equivalent to the BGK-LB model with the following equilibria [34]:

$$f_i^{(eq)} = \omega_i r \rho \left[ 1 + \frac{\mathbf{e}_i \cdot \mathbf{u}}{c_s^2} + \frac{(\mathbf{e}_i \cdot \mathbf{u})^2}{2\phi c_s^4} - \frac{|\mathbf{u}|^2}{2\phi c_s^2} \right], \quad (32)$$

where $\omega_0 = 4/9$, $\omega_{1\sim 4} = 1/9$, and $\omega_{5\sim 8} = 1/36$.

*3.2 D2Q5 LB equation for the temperature field*

The temperature field governed by Eq. (12) is solved by using the thermal axisymmetric LB model of Li et al. [32]. The evolution equation for the temperature field is given by

$$g_i(\mathbf{x} + \mathbf{e}_i \delta_t, t + \delta_t) - g_i(\mathbf{x}, t) = -\omega_g \left[ g_i(\mathbf{x}, t) - g_i^{(eq)}(\mathbf{x}, t) \right] + \delta_t \psi_i(\mathbf{x}, t), \quad (33)$$

where $g_i(\mathbf{x}, t)$ is the temperature distribution function, $g_i^{(eq)}(\mathbf{x}, t)$ is the equilibrium temperature distribution function, and $\psi_i(\mathbf{x}, t)$ is the source term. The relaxation parameter $\omega_g$ is given by $\omega_g = \left[ 1 + (e_{ir} \tau_T \delta_t / r) \right] / (\tau_T + 0.5)$ [32], in which $\tau_T$ is the dimensionless relaxation time related to $\alpha_e$. In the present study, the D2Q5 model is employed, of which the five discrete velocities are

$$\mathbf{e}_i = \begin{cases} (0,0), & i = 0 \\ \left(\cos\left[(i-1)\pi/2\right], \sin\left[(i-1)\pi/2\right]\right)c, & i = 1 \sim 4 \end{cases}. \quad (34)$$

The equilibrium temperature distribution function $g_i^{(eq)}$ is defined as

$$g_i^{(eq)} = \tilde{\omega}_i T \left( 1 + \frac{\mathbf{e}_i \cdot \mathbf{u}}{c_{sT}^2} \right), \quad (35)$$

where $c_{sT}$ is the sound speed of the D2Q5 model, and $\{\tilde{\omega}_i | i = 0, 1, \ldots, 4\}$ are weight coefficients given by

$$\tilde{\omega}_i = \begin{cases} \tilde{\omega}_0, & i = 0 \\ (1-\tilde{\omega}_0)/4, & i = 1 \sim 4 \end{cases}. \tag{36}$$

For the D2Q5 model, $c_{sT}^2 = \sum_i \omega_i e_{ir}^2 = \sum_i \omega_i e_{iz}^2$, and $0 < \tilde{\omega}_0 < 1$. In the present work, $\tilde{\omega}_0$ is set to be 3/5 with $c_{sT} = c/\sqrt{5} = 1/\sqrt{5}$.

The source term $\psi_i$ can be chosen as [32]

$$\psi_i = -\frac{u_r}{r} g_i^{(eq)}. \tag{37}$$

The macroscopic temperature $T$ is defined as

$$T = \sum_{i=0}^{4} g_i. \tag{38}$$

Through the Chapman-Enskog analysis [32] of the LB equation (33), the macroscopic temperature equation (12) can be recovered in the incompressible limit. The effective thermal diffusivity $\alpha_e$ is given by $\alpha_e = c_{sT}^2 \tau_T \delta_t$. The temperature field governed by the axisymmetric convection-diffusion equation (12) can also be solved by the MRT-LB model. Details of the MRT-LB model for axisymmetric convection-diffusion equation can be found in Ref. [33].

## 4. Numerical validation

In this section, numerical simulations of several typical axisymmetric thermal flows in porous media are carried out to validate the present MRT-LB model. The test problems include the thermally developing flow in a pipe filled with porous media, natural convection flow in a vertical annulus, natural convection flow in a vertical annulus filled with porous media, and natural convection flow in a vertical porous annulus with discrete heating. In simulations, we set $\rho_0 = 1$, $\delta_t = 1$, $\delta_r = \delta_z = 1$ $c = 1$, $J = 1$, and $\lambda = 1$. Unless otherwise stated, the nonequilibrium extrapolation scheme [47] is adopted to treat the velocity and thermal boundary conditions of $f_i$ and $g_i$. The relaxation rates $\{s_i | 0 < s_i < 2\}$ are chosen as follows: $s_\rho = s_j = 1$, $s_e = s_\varepsilon = 1.1$, $s_q = 1.2$, and $s_\upsilon = 1/\tau_\upsilon$. For natural convection problems in porous annulus, the dimensionless relaxation time $\tau_\upsilon$ ($\tau_\upsilon = 1/s_\upsilon$) can be fully determined in terms of $Pr$, $Ra$ and $Ma$ [15, 48], where $Ma = U/c_s = \sqrt{3}U$ is the Mach number ($U = \sqrt{\beta g \Delta T L}$ is the characteristic velocity). The dimensionless relaxation times $\tau_\upsilon$ and $\tau_T$ can be determined as

$$\tau_\upsilon = 0.5 + \frac{MaJL\sqrt{Pr}}{c_s \delta_t \sqrt{Ra}}, \quad \tau_T = \frac{c_s^2 \lambda (\tau_\upsilon - 0.5)}{c_{sT}^2 JPr}, \tag{39}$$

where $c_s^2 = 1/3$, $c_{sT}^2 = 1/5$, and $Ma$ is set to be $0.1$ in the present work.

*4.1 Thermally developing flow in a pipe filled with porous media*

In this subsection, we apply the present LB model to study the thermally developing flow in a pipe filled with fluid-saturated porous media, which has been numerically investigated in Ref. [6]. The computational domain and boundary conditions of this problem are sketched in Fig. 1. $H$ and $D$ are the length and diameter of the pipe, respectively. At the inlet ($z=0$), $u_r = 0$, $u_z = U_{in}$, and $T = T_{in}$. At the outlet ($z = H$), the gradients of the velocity and temperature in the $z$ direction are set to zero (fully developed flow). For $0 < z < H$ and $r = r_i$ ($r_i = D/2$ is the inner radius of the pipe), no slip condition is imposed with $T = T_w$. The local Nusselt numbers along the pipe wall is defined as [6]

$$Nu = D(\partial_r T)_w / (T_w - T_b), \tag{40}$$

where $T_b$ is the bulk temperature

$$T_b = \int_0^{D/2} rT dr / \int_0^{D/2} r dr. \tag{41}$$

In simulations, we set $Pr = 0.7$, $\phi = 1$, $F_\phi = 0$, and $Re = LU_{in}/\upsilon = 100$ (the characteristic length $L = r_i = D/2$). A uniform grid $N_r \times N_z = 81 \times 648$ is employed, corresponding to an aspect ratio $A = H/r_i = 16$. The local Nusselt numbers along the pipe wall for different Darcy numbers are shown in Fig. 2. From the figure we can observe that the Nusselt number increases as the Darcy number decreases. The heat transfer effect can be enhanced by inserting porous material into the pipe.

On the other hand, the influence of the Darcy number on the pressure drop is significant when the pipe is fully filled with porous material. As the Darcy number decreases, the flow resistance increases. For more details on this topic, readers are referred to Ref. [6]. The Nusselt numbers for different Darcy numbers in the fully developed flow region are measured and included in Table 1. The numerical results given by Mohamad [6] using the finite volume method are also included in Table 1 for comparison. As shown in the table, the present results are in good agreement with those reported in the literature.

*4.2 Natural convection flow in a vertical annulus*

It is noted that as $\phi \to 1$ and $Da \to \infty$, the present LB model reduces to a LB model for axisymmetric thermal flows without porous media. In this subsection, we will test the present LB model by simulating the natural convection flow in a vertical annulus between two coaxial vertical cylinders without porous media, which has been numerically studied by many researchers [32, 33, 49, 50]. The configuration of the problem is sketched in Fig. 3. The inner cylinder wall (located at radius $r = r_i$) and outer cylinder wall (located at radius $r = r_o$) are maintained at constant but different temperatures $T_i$ and $T_o$ ($T_i > T_o$), respectively, while the horizontal walls are adiabatic. The radius ratio $R^* = r_o/r_i$ and aspect ratio $A = H/L$ are both set to be 2. The average Nusselt numbers on the inner cylinder wall ($Nu_i$) and outer cylinder wall ($Nu_o$) are defined as [32, 50]

$$Nu_{i,o} = -\frac{1}{H\Delta T} r_{i,o} \int_0^H (\partial_r T)_{i,o} \, dz, \qquad (42)$$

where $\Delta T = (T_i - T_o)$ is the temperature difference, $T_0 = (T_i + T_o)/2$ is the reference temperature, and $L = r_o - r_i$ is the characteristic length.

In simulations, we set $Pr = 0.7$, $\phi = 1$, and $Da = 10^8$. Numerical simulations are carried out for $Ra = 10^3$, $10^4$, and $10^5$ based on a $N_r \times N_z = 100 \times 200$ uniform grid. The streamlines and isotherms for different $Ra$ are illustrated in Fig. 4. The present results are in good agreement with those reported in previous studies [32, 33, 49, 50]. To quantify the results, the average Nusselt numbers on the inner cylinder wall ($Nu_i$) predicted by the present LB model are included in Table 2. The published data obtained by different numerical methods in Refs. [32, 33, 49, 50] are also listed in Table 2 for comparison. As shown in Table 2, the present numerical results agree well with the results reported in previous studies.

*4.3 Natural convection flow in a vertical annulus filled with porous media*

Natural convection heat transfer in a vertical porous annulus between two coaxial vertical cylinders has been investigated both experimentally and numerically by many researchers [3, 4, 8, 34]. In this subsection, we will validate the present LB model by simulating such flows. The configuration of the problem is the same as that sketched in Fig. 3, but the annulus is filled with fluid-saturated porous media. The numerical results predicted by the present LB model are validated against the numerical solutions of Prasad et al. [3, 4]. The aspect ratio $A$ and radius ratio $R^*$ of the annulus are set to be $1$ and $5.338$, respectively.

In simulations, we set $Pr = 1$, $\phi = 0.3698$, and $Da = 1.66 \times 10^{-6}$. Numerical simulations are carried out for $Ra^* = 200$, $500$, and $2000$ based on a $N_r \times N_z = 300 \times 300$ uniform grid. Here, $Ra^*$ is the Darcy-Rayleigh number defined as $Ra^* = RaDa$. The streamlines and isotherms predicted by the present LB model are shown in Fig. 5. From the figure it can be seen that as $Ra^*$ increases, the core of the flow field shifts towards the top wall of the annulus. For a larger $Ra^*$, the temperature gradient near the left hot wall is larger and the isotherms shift towards the left vertical cylinder wall. These observations agree well with those reported in Ref. [3]. In order to examine the behavior of the heat transfer inside the porous annulus, Fig. 6 illustrates the temperature profiles at four different heights ($z/H = 0.25$, $0.5$, $0.75$, $1$) of the annulus for different Darcy-Rayleigh numbers. The temperature profiles in the figure clearly indicate that, as the Darcy-Rayleigh number increases, the temperature at the center of the annulus decreases. The numerical results given by Prasad et al. [3, 4] are also included in Fig. 6 for comparison. As shown, the numerical results obtained by the present LB model agree well with the available results in previous studies.

*4.4 Natural convection flow in a vertical porous annulus with discrete heating*

In this subsection, the numerical results predicted by the present LB model for natural convection flow in a vertical porous annulus with an isoflux discrete heater are presented. The computational domain and boundary conditions of the problem are sketched in Fig. 7. An isoflux discrete heater of length $l$ ($l/H = 0.4$) and strength $q'$ ($q' = -k\partial_r T$, in which $k$ is the thermal conductivity) is placed on the inner cylinder wall of the annulus, and the unheated parts of the inner cylinder wall are thermally insulated. The outer cylinder wall is maintained at a constant temperature $T_o$, while the bottom and top walls are thermally insulated. The distance between the bottom wall and the center of the heater is $h$ ($h/H = 0.5$). The aspect ratio $A$ and radius ratio $R^*$ of the annulus are set to be $1$ and $2$, respectively. The average Nusselt number $\overline{Nu}$ along the discrete heater is defined as [5]

$$\overline{Nu} = \frac{1}{l}\int_{h-0.5l}^{h+0.5l} Nu(z)\,dz, \qquad (43)$$

where $Nu(z)$ is the local Nusselt number

$$Nu(z) = \frac{q'L}{k(T_h - T_o)}, \qquad (44)$$

in which $T_h$ is the local temperature of the discrete heater, and $L = r_o - r_i$ is the characteristic length. The temperature difference $\Delta T = q'L/k$, and the reference temperature $T_0 = T_o$.

In simulations, we set $Ra = 10^7$, $Pr = 0.7$, $\phi = 0.9$, $F_\phi = 0$, $\Delta T = 10$, and $T_o = 1$. Numerical simulations are carried out for $Da = 10^{-2}$, $10^{-4}$, and $10^{-6}$ based on a $N_r \times N_z = 200 \times 200$ uniform grid. The streamlines and isotherms for different Darcy numbers are plotted in Fig. 8, from which we can observe that the fluid flow and heat transfer inside the annulus strongly depend on $Da$. At low Darcy number $Da = 10^{-6}$, the strength of the flow field is weak and a weak-convection structure can be observed from the isotherms. As $Da$ increases to $10^{-2}$, more convective mixing occurs inside the annulus, and the main vortex moves towards the cold vertical cylinder wall. Moreover, it can be observed that the local temperature inside the annulus decreases rapidly as the Darcy number increases to $10^{-2}$. To quantify the results, the average Nusselt numbers along the discrete heater and the maximum values of the stream function are measured and listed in Table 3 together with the results from Ref. [5]. As shown, our results agree well with those reported in previous studies.

## 5. Conclusions

In this paper, we have presented an axisymmetric MRT-LB model for simulating incompressible axisymmetric thermal flows in porous media at the REV scale. In the model, a D2Q9 MRT-LB equation is proposed to solve the flow field in addition to the D2Q5 LB equation for the temperature field. The source terms of the present MRT-LB model are simple and contain no velocity and temperature gradient terms. The equilibrium moments are modified to account for the porosity of the porous media, and the linear and nonlinear drag forces of the porous matrix are incorporated into the model by adding a source term into the MRT-LB equation in the moment space. Through the Chapman-Enskog analysis of the MRT-LB equation in the moment space, the generalized axisymmetric momentum equation can be correctly derived in the incompressible limit. The effectiveness and accuracy of the present LB model is demonstrated by numerical simulations of several typical axisymmetric thermal problems in porous media. It is found that the present results agree well with the data reported in previous studies.

**Acknowledgements**

This work was supported by the National Key Basic Research Program of China (973 Program) (2013CB228304).

**References**

[1] Nield DA, Bejan A. Convection in Porous Media. New York: Springer; 2006.

[2] Vafai K. Handbook of Porous Media. New York: Taylor & Francis; 2005.

[3] Prasad V, Kulacki FA. Natural convection in porous media bounded by short concentric vertical cylinders. J Heat Transfer 1985;107:147-154.

[4] Prasad V, Kulacki FA, Keyhani M. Natural convection in porous media. J Fluid Mech 1985;150:89-119.

[5] Sankar M, Park Y, Lopez JM, Do Y. Numerical study of natural convection in a vertical porous annulus with discrete heating. Int J Heat Mass Transfer 2011;54:1493-1505.

[6] Mohamad AA. Heat transfer enhancement in heat exchangers fitted with porous media. Part I: constant wall temperature. Int J Therm Sci 2003;42:385-395.

[7] Teamah MA, El-Maghlany WM, Khairat Dawood MM. Numerical simulation of laminar forced convection in horizontal pipe partially or completely filled with porous material. Int J Therm Sci 2011;50:1512-1522.

[8] Arpino F, Carotenuto A, Massarotti N, Mauro A. New solutions for axial flow convection in porous and partly porous cylindrical domains. Int J Heat Mass Transfer 2013;57:155-170.

[9] Frisch U, Hasslacher B, Pomeau Y. Lattice-gas automata for the Navier-Stokes equation. Phys Rev Lett 1986;56:1505-1508.

[10] Chen S, Doolen GD. Lattice Boltzmann method for fluid flows. Annu Rev Fluid Mech

1998;30:329-364.

[11] Succi S. The Lattice Boltzmann Equation for Fluid Dynamics and Beyond. Oxford: Clarendon Press; 2001.

[12] Kang Q, Zhang D, Chen S. Unified lattice Boltzmann method for flow in multiscale porous media. Phys Rev E 2002;66:056307.

[13] Li Q, He YL, Wang Y, Tao WQ. Coupled double-distribution-function lattice Boltzmann method for the compressible Navier-Stokes equations. Phys Rev E 2007;76:056705.

[14] Succi S. Lattice Boltzmann across scales: from turbulence to DNA translocation. Eur Phys J B 2008;64:471-479.

[15] He YL, Wang Y, Li Q. Lattice Boltzmann Method: Theory and Applications. Beijing: Science Press; 2009.

[16] Feng Y, Sagaut P, Tao W. A three dimensional lattice model for thermal compressible flow on standard lattices. J Comput Phys 2015; 303: 514-529.

[17] He YL, Liu Q, Li Q. Three-dimensional finite-difference lattice Boltzmann model and its application to inviscid compressible flows with shock waves. Physica A 2013;392:4884-4896.

[18] Wang J, Wang D, Lallemand P, Luo LS. Lattice Boltzmann simulations of thermal convective flows in two dimensions. Comput Math Appl 2013;66:262-286.

[19] Halliday I, Hammond LA, Care CM, Good K, Stevens A. Lattice Boltzmann equation hydrodynamics. Phys Rev E 2001;64:011208.

[20] Lee TS, Huang H, Shu C. An axisymmetric incompressible lattice Boltzmann model for pipe flow. Int J Mod Phys C 2006;17:645-661.

[21] Reis T, Phillips TN. Modified lattice Boltzmann model for axisymmetric flows. Phys Rev E

2007;75:056703.

[22] Reis T, Phillips TN. Numerical validation of a consistent axisymmetric lattice Boltzmann model. Phys Rev E 2008;77:026703.

[23] Zhou JG. Axisymmetric lattice Boltzmann method. Phys Rev E 2008;78:036701.

[24] Guo Z, Han H, Shi B, Zheng C. Theory of the lattice Boltzmann equation: lattice Boltzmann model for axisymmetric flows. Phys Rev E 2009;79:046708.

[25] Wang L, Guo Z, Zheng C. Multi-relaxation-time lattice Boltzmann model for axisymmetric flows. Comput Fluids 2010;39:1542-1548.

[26] Li Q, He YL, Tang GH, Tao WQ. Improved axisymmetric lattice Boltzmann scheme. Phys Rev E 2010;81:056707.

[27] Tang GH, Li XF, Tao WQ. Microannular electro-osmotic flow with the axisymmetric lattice Boltzmann method. J Appl Phys 2010;108:114903.

[28] Zhou JG. Axisymmetric lattice Boltzmann method revised. Phys Rev E 2011;84:036704.

[29] Wang Y, Shu C, Teo CJ. A fractional step axisymmetric lattice Boltzmann flux solver for incompressible swirling and rotating flows. Comput Fluids 2014;96:204-214.

[30] Peng Y, Shu C, Chew YT, Qiu J. Numerical investigation of flows in Czochralski crystal growth by an axisymmetric lattice Boltzmann method. J Comput Phys 2003;186:295-307.

[31] Huang H, Lee TS, Shu C. Hybrid lattice Boltzmann finite-difference simulation of axisymmetric swirling and rotating flows. Int J Numer Meth Fluids 2007; 53:1707-1726.

[32] Li Q, He YL, Tang GH, Tao WQ. Lattice Boltzmann model for axisymmetric thermal flows. Phys Rev E 2009;80:037702.

[33] Li L, Mei R, Klausner JF. Multiple-relaxation-time lattice Boltzmann model for the axisymmetric

convection diffusion equation. Int J Heat Mass Transfer 2013;67:338-351.

[34] Rong F, Guo Z, Chai Z, Shi B. A lattice Boltzmann model for axisymmetric thermal flows through porous media. Int J Heat Mass Transfer 2010;53:5519-5527.

[35] d'Humières D. Generalized lattice-Boltzmann equations, in rarefied gas dynamics: theory and simulations. Prog Astronaut Aeronaut 1992;159:450-458.

[36] Lallemand P, Luo LS. Theory of the lattice Boltzmann method: Dispersion, dissipation, isotropy, Galilean invariance, and stability. Phys Rev E 2000;61:6546-6262.

[37] Qian YH, d'Humières D, Lallemand P. Lattice BGK models for Navier-Stokes equation. Europhys Lett 1992;17:479-484.

[38] Liu Q, He YL, Li Q, Tao WQ. A multiple-relaxation-time lattice Boltzmann model for convection heat transfer in porous media. Int J Heat Mass Transfer 2014;73:761-775.

[39] Nithiarasu P, Seetharamu KN, Sundararajan T. Non-Darcy double-diffusive natural convection in axisymmetric fluid saturated porous cavities. Heat Mass Transfer 1997;32:427-433.

[40] Hsu CT, Cheng P. Thermal dispersion in a porous medium. Int J Heat Mass Transfer 1990;33:1587-1597.

[41] Guo Z, Zhao TS. Lattice Boltzmann model for incompressible flows through porous media. Phys Rev E 2002;66:036304.

[42] Ergun S. Fluid flow through packed columns. Chem Eng Prog 1952;48:89-94.

[43] Vafai K. Convective flow and heat transfer in variable-porosity media. J Fluid Mech 1984;147:233-259.

[44] McCracken ME, Abraham J. Multiple-relaxation-time lattice-Boltzmann model for multiphase flow. Phys Rev E 2005;71:036701.

[45] Li Q, Luo KH, He YL, Gao YJ, Tao WQ. Coupling lattice Boltzmann model for simulation of thermal flows on standard lattices. Phys Rev E 2012;85:016710.

[46] Li L, Mei R, Klausner JF. Boundary conditions for thermal lattice Boltzmann equation method. J Comput Phys 2013;237:366-395.

[47] Guo ZL, Zheng CG, Shi BC. Non-equilibrium extrapolation method for velocity and pressure boundary conditions in the lattice Boltzmann method. Chin Phys 2002;11:366-374.

[48] Li Q, He YL, Wang Y, Tang GH. An improved thermal lattice Boltzmann model for flows without viscous heat dissipation and compression work. Int J Mod Phys C 2008;19:125-150.

[49] Kumar R, Kalam MA. Laminar thermal convection between vertical coaxial isothermal cylinders. Int J Heat Mass Transfer 1991;34:513-524.

[50] Venkatachalappa M, Sankar M, Natarajan AA. Natural convection in an annulus between two rotating vertical cylinders. Acta Mechanica 2001;147:173-196.

[51] Chapman S, Cowling TG. The Mathematical Theory of Non-Uniform Gases. London: Cambridge University Press; 1970.

[52] Chai Z, Zhao TS. Effect of the forcing term in the multiple-relaxation-time lattice Boltzmann equation on the shear stress or the strain rate tensor. Phys Rev E 2012;86:016705.

**Appendix A: Chapman-Enskog analysis of the axisymmetric D2Q9 MRT-LB model**

The Chapman-Enskog expansion method [35, 44, 51, 52] is adopted to derive the generalized axisymmetric Navier-Stokes equations (10) and (11) from the present MRT-LB model. To this end, the following expansions in time and space are introduced [52]:

$$f_i\left(\mathbf{x}+\mathbf{e}_i\delta_t,\, t+\delta_t\right) = \sum_{i=0}^{\infty} \frac{\epsilon^n}{n!}\left(\partial_t + \mathbf{e}_i\cdot\nabla\right)^n f_i\left(\mathbf{x},\, t\right), \qquad (A.1a)$$

$$f_i = f_i^{(0)} + \epsilon f_i^{(1)} + \epsilon^2 f_i^{(2)} + \cdots, \qquad (A.1b)$$

$$\partial_t = \epsilon \partial_{t_1} + \epsilon^2 \partial_{t_2}, \quad \partial_j = \epsilon \partial_{j1}, \quad \mathbf{S} = \epsilon \mathbf{S}_1, \quad \tilde{\mathbf{F}} = \epsilon \tilde{\mathbf{F}}_1 \tag{A.1c}$$

where $\epsilon = \delta_t$ is a small expansion parameter, $\mathbf{S}_1 = (S_{01}, S_{11}, \ldots, S_{81})^{\mathrm{T}}$, $\tilde{\mathbf{F}}_1 = (\tilde{F}_{r1}, \tilde{F}_{z1})$. With the above expansions, we can derive the following equations as consecutive orders of the parameter $\epsilon$:

$$\epsilon^0: \quad f_i^{(0)} = f_i^{(eq)}, \tag{A.2a}$$

$$\epsilon^1: \quad D_{1i} f_i^{(0)} = -\frac{1}{\delta_t} \left( \mathbf{M}^{-1} \boldsymbol{\Lambda} \mathbf{M} \right)_{i\beta} f_\beta^{(1)} + \left[ \mathbf{M}^{-1} \left( \mathbf{I} - \frac{\boldsymbol{\Lambda}}{2} \right) \mathbf{M} \right]_{i\beta} \tilde{S}_{\beta 1}, \tag{A.2b}$$

$$\epsilon^2: \quad \partial_{t_2} f_i^{(0)} + D_{1i} f_i^{(1)} + \frac{\delta_t}{2} D_{1i}^2 f_i^{(0)} = -\frac{1}{\delta_t} \left( \mathbf{M}^{-1} \boldsymbol{\Lambda} \mathbf{M} \right)_{i\beta} f_\beta^{(2)}, \tag{A.2c}$$

where $D_{1i} = \partial_{t_1} + \mathbf{e}_i \cdot \nabla_1 = \partial_{t_1} + e_{ij} \partial_{j1}$ ($j = r, z$), $\tilde{\mathbf{S}}_1 = (\tilde{S}_{01}, \tilde{S}_{11}, \ldots, \tilde{S}_{81}) = \mathbf{M}^{-1} \mathbf{S}_1$. In the moment space, the above equations can be rewritten as follows:

$$\epsilon^0: \quad \mathbf{m}^{(0)} = \mathbf{m}^{(eq)}, \tag{A.3a}$$

$$\epsilon^1: \quad \tilde{\mathbf{D}}_1 \mathbf{m}^{(0)} = -\boldsymbol{\Lambda}' \mathbf{m}^{(1)} + \left( \mathbf{I} - \frac{\boldsymbol{\Lambda}}{2} \right) \mathbf{S}_1, \tag{A.3b}$$

$$\epsilon^2: \quad \partial_{t_2} \mathbf{m}^{(0)} + \tilde{\mathbf{D}}_1 \left( \mathbf{I} - \frac{\boldsymbol{\Lambda}}{2} \right) \mathbf{m}^{(1)} + \frac{\delta_t}{2} \tilde{\mathbf{D}}_1 \left( \mathbf{I} - \frac{\boldsymbol{\Lambda}}{2} \right) \mathbf{S}_1 = -\boldsymbol{\Lambda}' \mathbf{m}^{(2)}, \tag{A.3c}$$

where $\tilde{\mathbf{D}}_1 = \mathbf{M} \mathbf{D}_1 \mathbf{M}^{-1} = \partial_{t_1} \mathbf{I} + \mathbf{C}_j \partial_{j1}$, $\mathbf{C}_j = \mathbf{M}(e_{ij} \mathbf{I}) \mathbf{M}^{-1}$, $\boldsymbol{\Lambda}' = \boldsymbol{\Lambda}/\delta_t$, and

$$\mathbf{m}^{(1)} = \left( 0, e^{()}, \varepsilon^{()1}, -\delta_t \rho \tilde{F}_{r1}/2, q_r^{(),1}, -\delta_t \rho \tilde{F}_{z1}/2, q_z^{()}, p_{rr}^{()}, p_{rz}^{()} \right)^{\mathrm{T}}. \tag{A.4}$$

$\mathbf{C}_j$ can be given explicitly by

$$\mathbf{C}_r = \begin{pmatrix} 0 & 0 & 0 & 1 & 0 & 0 & 0 & 0 & 0 \\ 0 & 0 & 0 & 1 & 1 & 0 & 0 & 0 & 0 \\ 0 & 0 & 0 & 0 & 1 & 0 & 0 & 0 & 0 \\ \frac{2}{3} & \frac{1}{6} & 0 & 0 & 0 & 0 & 0 & \frac{1}{2} & 0 \\ 0 & \frac{1}{3} & \frac{1}{3} & 0 & 0 & 0 & 0 & -1 & 0 \\ 0 & 0 & 0 & 0 & 0 & 0 & 0 & 0 & 1 \\ 0 & 0 & 0 & 0 & 0 & 0 & 0 & 0 & 1 \\ 0 & 0 & 0 & \frac{1}{3} & -\frac{1}{3} & 0 & 0 & 0 & 0 \\ 0 & 0 & 0 & 0 & 0 & \frac{2}{3} & \frac{1}{3} & 0 & 0 \end{pmatrix}, \mathbf{C}_z = \begin{pmatrix} 0 & 0 & 0 & 0 & 0 & 1 & 0 & 0 & 0 \\ 0 & 0 & 0 & 0 & 0 & 1 & 1 & 0 & 0 \\ 0 & 0 & 0 & 0 & 0 & 0 & 1 & 0 & 0 \\ 0 & 0 & 0 & 0 & 0 & 0 & 0 & 0 & 1 \\ 0 & 0 & 0 & 0 & 0 & 0 & 0 & 0 & 1 \\ \frac{2}{3} & \frac{1}{6} & 0 & 0 & 0 & 0 & 0 & -\frac{1}{2} & 0 \\ 0 & \frac{1}{3} & \frac{1}{3} & 0 & 0 & 0 & 0 & 1 & 0 \\ 0 & 0 & 0 & 0 & 0 & -\frac{1}{3} & \frac{1}{3} & 0 & 0 \\ 0 & 0 & 0 & \frac{2}{3} & \frac{1}{3} & 0 & 0 & 0 & 0 \end{pmatrix}$$

From Eq. (A.3b), the following equations in the moment space at the $t_1$ time scale can be obtained:

$$r \partial_{t_1} \rho + \partial_{r1} (r \rho u_r) + r \partial_{z1} (\rho u_z) = 0, \tag{A.5a}$$

$$r\partial_{t_1}\left(-2\rho+\frac{3\rho|\mathbf{u}|^2}{\phi}\right)=-s_1' re^{(1)}+\left(1-\frac{s_1}{2}\right)S_{11}, \tag{A.5b}$$

$$r\partial_{t_1}\left(\rho-\frac{3\rho|\mathbf{u}|^2}{\phi}\right)+\partial_{r1}\left(-r\rho u_r\right)+r\partial_{z1}\left(-\rho u_z\right)=-s_2' r\varepsilon^{(1)}+\left(1-\frac{s_2}{2}\right)S_{21}, \tag{A.5c}$$

$$r\partial_{t_1}\left(\rho u_r\right)+\partial_{r1}\left(\frac{r\rho}{3}+\frac{r\rho u_r^2}{\phi}\right)+r\partial_{z1}\left(\frac{\rho u_r u_z}{\phi}\right)=\frac{\delta_t}{2}s_3' r\rho\tilde{F}_{r1}+\left(1-\frac{s_3}{2}\right)S_{31}, \tag{A.5d}$$

$$r\partial_{t_1}\left(-\rho u_r\right)+\partial_{r1}\left[-\frac{r\rho}{3}-\frac{r\rho\left(u_r^2-u_z^2\right)}{\phi}\right]+r\partial_{z1}\left(\frac{\rho u_r u_z}{\phi}\right)=-s_4' rq_r^{(1)}+\left(1-\frac{s_4}{2}\right)S_{41}, \tag{A.5e}$$

$$r\partial_{t_1}\left(\rho u_z\right)+\partial_{r1}\left(\frac{r\rho u_r u_z}{\phi}\right)+r\partial_{z1}\left(\frac{\rho}{3}+\frac{\rho u_z^2}{\phi}\right)=\frac{\delta_t}{2}s_5' r\rho\tilde{F}_{z1}+\left(1-\frac{s_5}{2}\right)S_{51}, \tag{A.5f}$$

$$r\partial_{t_1}\left(-\rho u_z\right)+\partial_{r1}\left(\frac{r\rho u_r u_z}{\phi}\right)+r\partial_{z1}\left[-\frac{\rho}{3}+\frac{\rho\left(u_r^2-u_z^2\right)}{\phi}\right]=-s_6' rq_z^{(1)}+\left(1-\frac{s_6}{2}\right)S_{61}, \tag{A.5g}$$

$$r\partial_{t_1}\left[\frac{\rho\left(u_r^2-u_z^2\right)}{\phi}\right]+\partial_{r1}\left(\frac{2r\rho u_r}{3}\right)+r\partial_{z1}\left(-\frac{2\rho u_z}{3}\right)=-s_7' rp_{rr}^{(1)}+\left(1-\frac{s_7}{2}\right)S_{71}, \tag{A.5h}$$

$$r\partial_{t_1}\left(\frac{\rho u_r u_z}{\phi}\right)+\partial_{r1}\left(\frac{r\rho u_z}{3}\right)+r\partial_{z1}\left(\frac{\rho u_r}{3}\right)=-s_8' rp_{rz}^{(1)}+\left(1-\frac{s_8}{2}\right)S_{81}. \tag{A.5i}$$

From Eq. (A.3c), the following equations at the $t_2$ time scale corresponding to the conservative variables $\rho$, $\rho u_r$, and $\rho u_z$ can be obtained:

$$\partial_{t_2}\rho=0, \tag{A.6a}$$

$$r\partial_{t_2}\left(\rho u_r\right)-\frac{\delta_t}{2}r\partial_{t_1}\left[\left(1-\frac{s_3}{2}\right)\rho\tilde{F}_{r1}\right]+\partial_{r1}\left[\frac{1}{6}\left(1-\frac{s_1}{2}\right)re^{(1)}+\frac{1}{2}\left(1-\frac{s_7}{2}\right)rp_{rr}^{(1)}\right]+r\partial_{z1}\left[\left(1-\frac{s_8}{2}\right)p_{rz}^{(1)}\right]$$

$$+\frac{\delta_t}{2}\left\{\partial_{t_1}\left[\left(1-\frac{s_3}{2}\right)S_{31}\right]+\partial_{r1}\left[\frac{1}{6}\left(1-\frac{s_1}{2}\right)S_{11}+\frac{1}{2}\left(1-\frac{s_7}{2}\right)S_{71}\right]+\partial_{z1}\left[\left(1-\frac{s_8}{2}\right)S_{81}\right]\right\}=0, \tag{A.6b}$$

$$r\partial_{t_2}\left(\rho u_z\right)-\frac{\delta_t}{2}r\partial_{t_1}\left[\left(1-\frac{s_5}{2}\right)\rho\tilde{F}_{z1}\right]+\partial_{r1}\left[\left(1-\frac{s_8}{2}\right)rp_{rz}^{(1)}\right]+r\partial_{z1}\left[\frac{1}{6}\left(1-\frac{s_1}{2}\right)e^{(1)}-\frac{1}{2}\left(1-\frac{s_7}{2}\right)p_{rr}^{(1)}\right]$$

$$+\frac{\delta_t}{2}\left\{\partial_{t_1}\left[\left(1-\frac{s_5}{2}\right)S_{51}\right]+\partial_r\left[\left(1-\frac{s_8}{2}\right)S_{81}\right]+\partial_z\left[\frac{1}{6}\left(1-\frac{s_1}{2}\right)S_{11}-\frac{1}{2}\left(1-\frac{s_7}{2}\right)S_{71}\right]\right\}=0, \tag{A.6c}$$

Note that $e^{(1)}$, $p_{rr}^{(1)}$ and $p_{rz}^{(1)}$ in Eqs. (A.6b) and (A.6c) are unknowns to be determined. With the aid of Eqs. (A.5b), (A.5h) and (A.5i), we can get:

$$-s_1'e^{(1)} = \partial_{t_1}\left[-2\rho + \frac{3\rho(u_r^2+u_z^2)}{\phi}\right] - \frac{1}{r}\left(1-\frac{s_1}{2}\right)S_{11}, \quad (A.7a)$$

$$-s_7'p_{rr}^{(1)} = \partial_{t_1}\left[\frac{\rho(u_r^2-u_z^2)}{\phi}\right] + \frac{2}{3}\frac{1}{r}\partial_{r1}(r\rho u_r) - \frac{2}{3}\partial_{z1}(\rho u_z) - \frac{1}{r}\left(1-\frac{s_7}{2}\right)S_{71}, \quad (A.7b)$$

$$-s_8'p_{rz}^{(1)} = \partial_{t_1}\left(\frac{\rho u_r u_z}{\phi}\right) + \frac{1}{3}\frac{1}{r}\partial_{r1}(r\rho u_z) + \frac{1}{3}\partial_{z1}(\rho u_r) - \frac{1}{r}\left(1-\frac{s_8}{2}\right)S_{81}. \quad (A.7c)$$

Neglecting the terms of order $O(|\mathbf{u}|^3)$ and higher-order terms of the form $u_j\partial_k(u_k u_j)$, using Eqs. (A.5a), (A.5d) and (A.5f), we can obtain:

$$\partial_{t_1}\rho = -\rho\left[\partial_{r1}u_r + \partial_{z1}u_z + \left(\frac{u_r}{r}\right)_1\right], \quad (A.8a)$$

$$\partial_{t_1}\left(\frac{\rho u_r^2}{\phi}\right) = \frac{2\rho u_r \tilde{F}_{r1}}{\phi}, \quad (A.8b)$$

$$\partial_{t_1}\left(\frac{\rho u_z^2}{\phi}\right) = \frac{2\rho u_z \tilde{F}_{z1}}{\phi}, \quad (A.8c)$$

$$\partial_{t_1}\left(\frac{\rho u_r u_z}{\phi}\right) = \frac{\rho(u_r\tilde{F}_{z1}+u_z\tilde{F}_{r1})}{\phi}, \quad (A.8d)$$

where $u_r/r = \epsilon(u_r/r)_1$. With the above equations, we can obtain:

$$-s_e'e^{(1)} = 2\rho\left[\partial_{r1}u_r + \partial_{z1}u_z + \left(\frac{u_r}{r}\right)_1\right] + 3s_e\frac{\rho(u_r\tilde{F}_{r1}+u_z\tilde{F}_{z1})}{\phi}, \quad (A.9a)$$

$$-s_\upsilon'p_{rr}^{(1)} = \frac{2\rho}{3}(\partial_{r1}u_r - \partial_{z1}u_z) + s_\upsilon\frac{\rho(u_r\tilde{F}_{r1}-u_z\tilde{F}_{z1})}{\phi}, \quad (A.9b)$$

$$-s_\upsilon'p_{rz}^{(1)} = \frac{\rho}{3}(\partial_{r1}u_z + \partial_{z1}u_r) + \frac{s_\upsilon}{2}\frac{\rho(u_r\tilde{F}_{z1}+u_z\tilde{F}_{r1})}{\phi}. \quad (A.9c)$$

Substituting Eq. (A.9) into Eq. (A.6), the following equations at the $t_2$ time scale can be derived:

$$\partial_{t_2}\rho = 0, \quad (A.10a)$$

$$r\partial_{t_2}(\rho u_r) = \partial_{r1}\left[\rho r\upsilon_e(\partial_{r1}u_r - \partial_{z1}u_z) + \rho r\upsilon_B\left(\partial_{r1}u_r + \partial_{z1}u_z + \left(\frac{u_r}{r}\right)_1\right)\right]$$

$$+ r\partial_{z1}\left[\rho\upsilon_e(\partial_{r1}u_z + \partial_{z1}u_r)\right], \quad (A.10b)$$

$$r\partial_{t_2}(\rho u_z) = r\partial_{z1}\left[\rho\upsilon_e\left(\partial_{z1}u_z - \partial_{r1}u_r\right) + \rho\upsilon_B\left(\partial_{r1}u_r + \partial_{z1}u_z + \left(\frac{u_r}{r}\right)_1\right)\right]$$

$$+\partial_{r1}\left[\rho r\upsilon_e\left(\partial_{r1}u_z + \partial_{z1}u_r\right)\right], \tag{A.10c}$$

where $\upsilon_e$ and $\upsilon_B$ are the effective kinetic viscosity and bulk viscosity

$$\upsilon_e = c_s^2\left(\frac{1}{s_\upsilon} - \frac{1}{2}\right)\delta_t, \quad \upsilon_B = c_s^2\left(\frac{1}{s_e} - \frac{1}{2}\right)\delta_t. \tag{A.11}$$

Combining Eq. (A.10) ($t_2$ time scale) with Eq. (A.5) ($t_1$ time scale) ($\partial_t = \epsilon\partial_{t_1} + \epsilon^2\partial_{t_2}$), the generalized axisymmetric Navier-Stokes equations (10) and (11) can be obtained in the incompressible limit ($\rho = \rho_0 + \delta\rho \approx \rho_0$, where $\rho_0$ is the mean fluid density, $\delta\rho$ is the density fluctuation and is of the order $O(|\mathbf{u}|^2)$).

**Figure Captions**

Fig. 1. Computational domain and boundary conditions of the thermally developing flow in a pipe filled with porous media.

Fig. 2. Local Nusselt numbers along the pipe wall for different Darcy numbers.

Fig. 3. Computational domain and boundary conditions of the natural convection flow in a vertical annulus.

Fig. 4. Streamlines (a) and isotherms (b) for $Ra = 10^3$ (left), $Ra = 10^4$ (middle), and $Ra = 10^5$ (right).

Fig. 5. Streamlines (left) and isotherms (right) for different Darcy-Rayleigh numbers: (a) $Ra^* = 200$; (b) $Ra^* = 500$; (c) $Ra^* = 2000$.

Fig. 6. Temperature profiles at four different heights of the annulus for different Darcy-Rayleigh numbers: (a) $Ra^* = 200$; (b) $Ra^* = 500$; (c) $Ra^* = 2000$.

Fig. 7. Computational domain and boundary conditions of the natural convection flow in a vertical porous annulus with discrete heating.

Fig. 8. Streamlines (left) and isotherms (right) for different Darcy numbers: (a) $Da = 10^{-6}$; (b) $Da = 10^{-4}$; (c) $Da = 10^{-2}$.

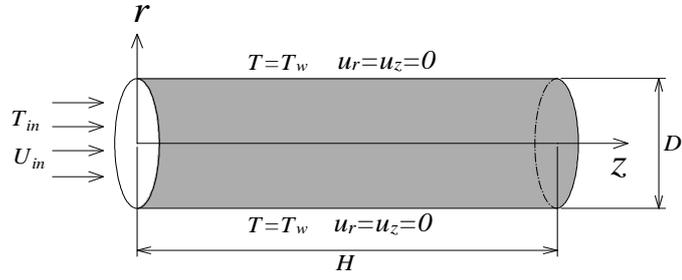

Fig. 1. Computational domain and boundary conditions of the thermally developing flow in a pipe filled with porous media.

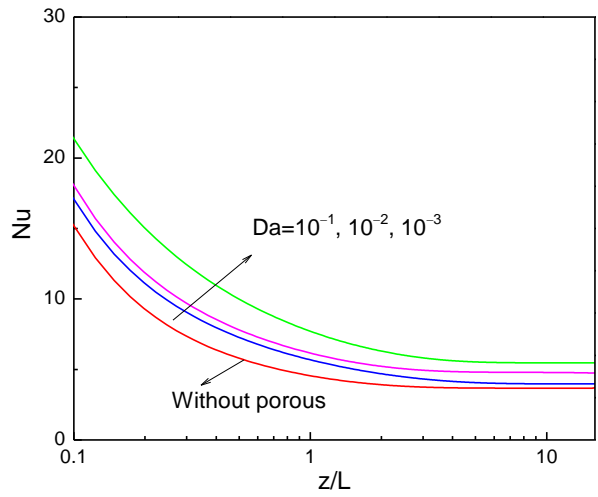

Fig. 2. Local Nusselt numbers along the pipe wall for different Darcy numbers.

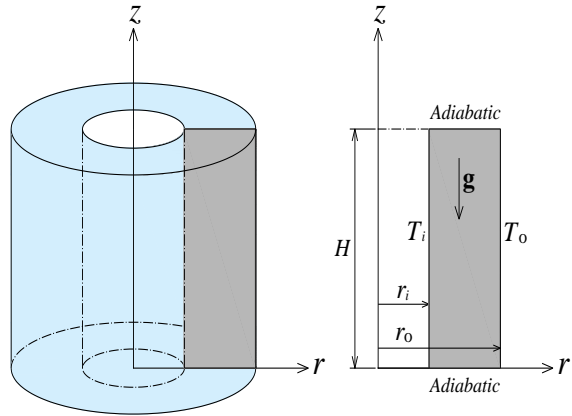

Fig. 3. Computational domain and boundary conditions of the natural convection flow in a vertical annulus.

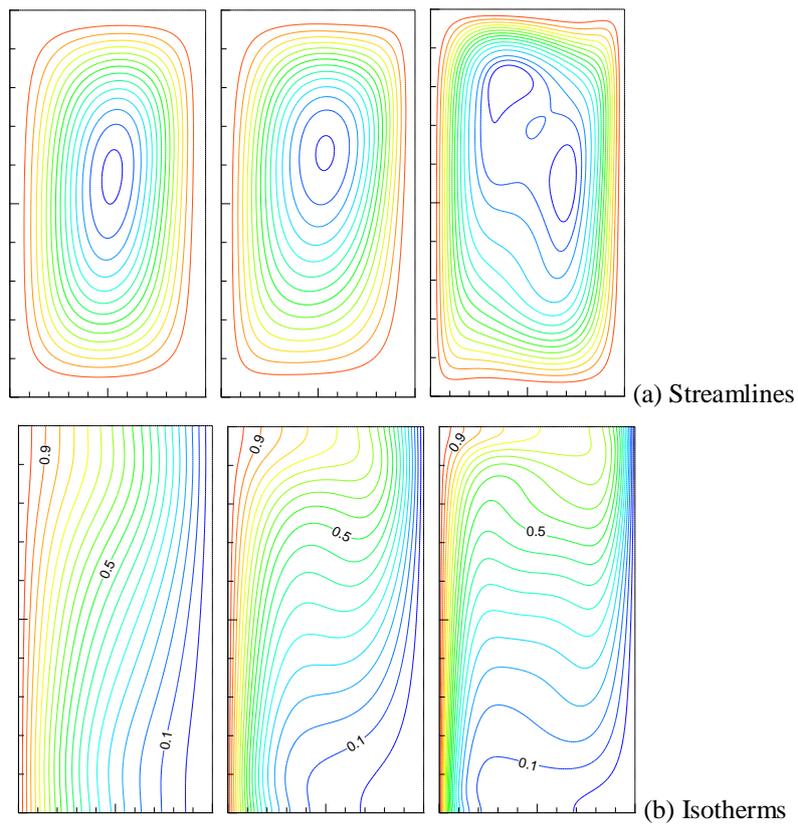

Fig. 4. Streamlines (a) and isotherms (b) for $Ra = 10^3$ (left), $Ra = 10^4$ (middle), and $Ra = 10^5$ (right).

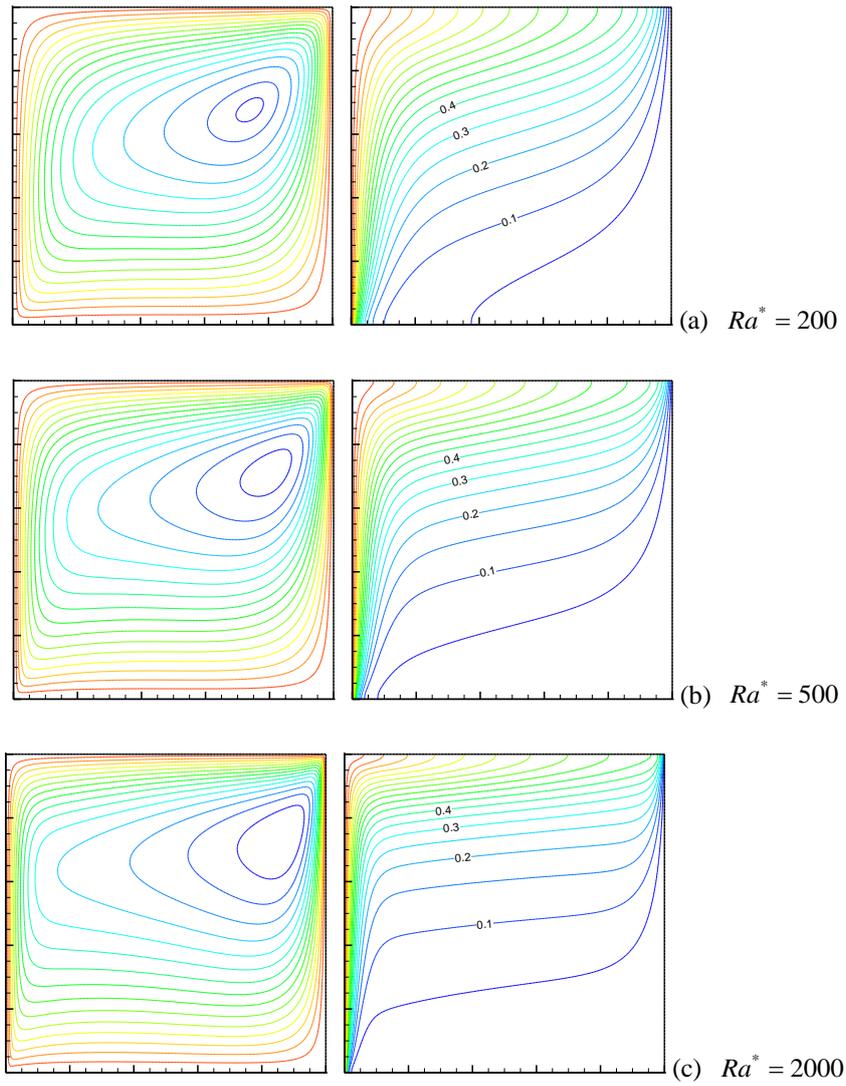

Fig. 5. Streamlines (left) and isotherms (right) for different Darcy-Rayleigh numbers: (a) $Ra^* = 200$; (b) $Ra^* = 500$; (c) $Ra^* = 2000$.

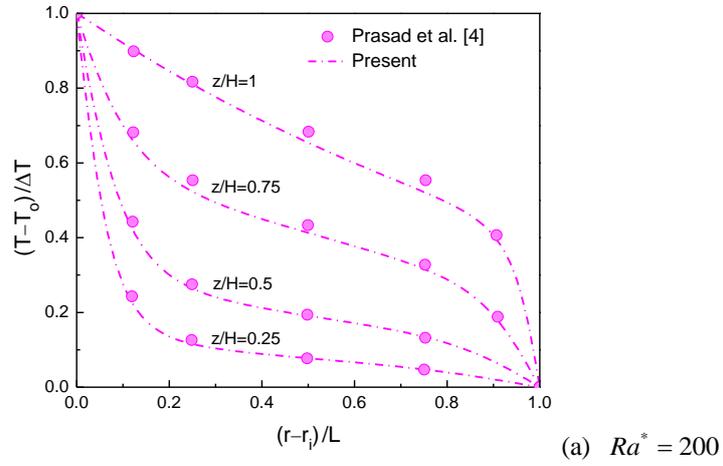

(a) $Ra^* = 200$

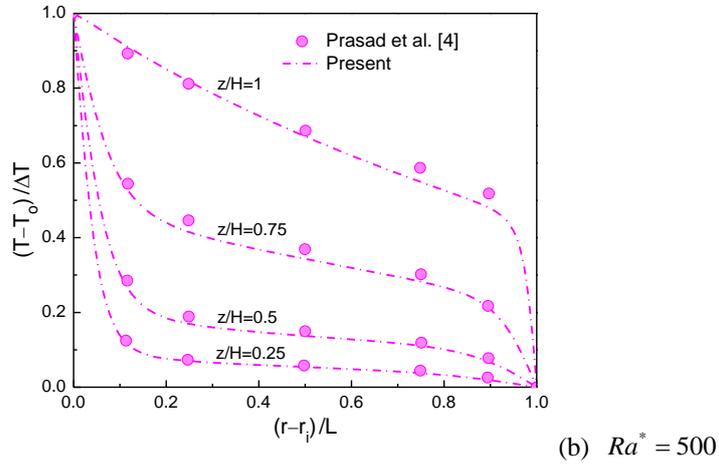

(b) $Ra^* = 500$

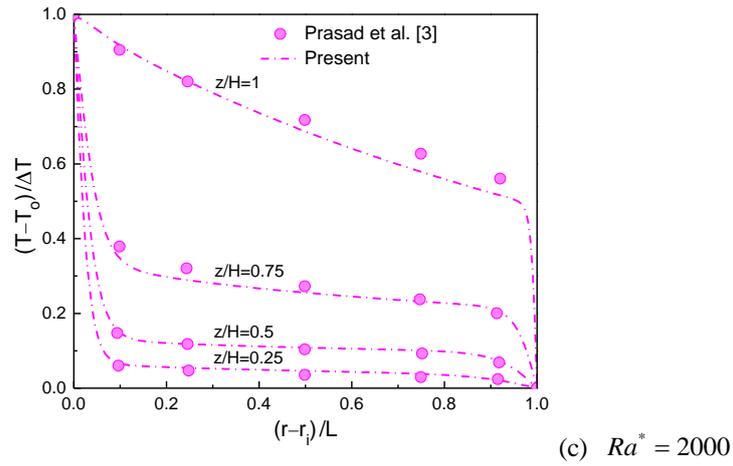

(c) $Ra^* = 2000$

Fig. 6. Temperature profiles at four different heights of the annulus for different Darcy-Rayleigh numbers: (a) $Ra^* = 200$; (b) $Ra^* = 500$; (c) $Ra^* = 2000$.

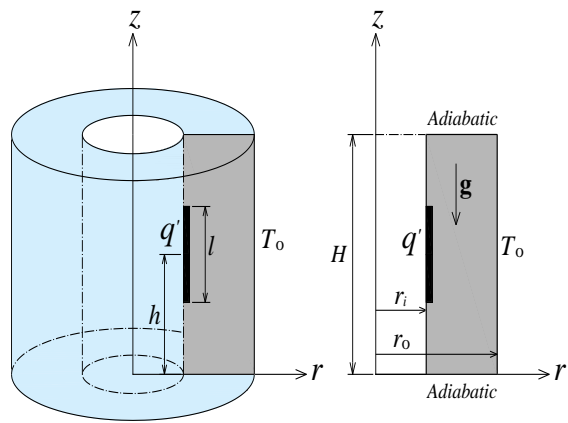

Fig. 7. Computational domain and boundary conditions of the natural convection flow in a vertical porous annulus with discrete heating.

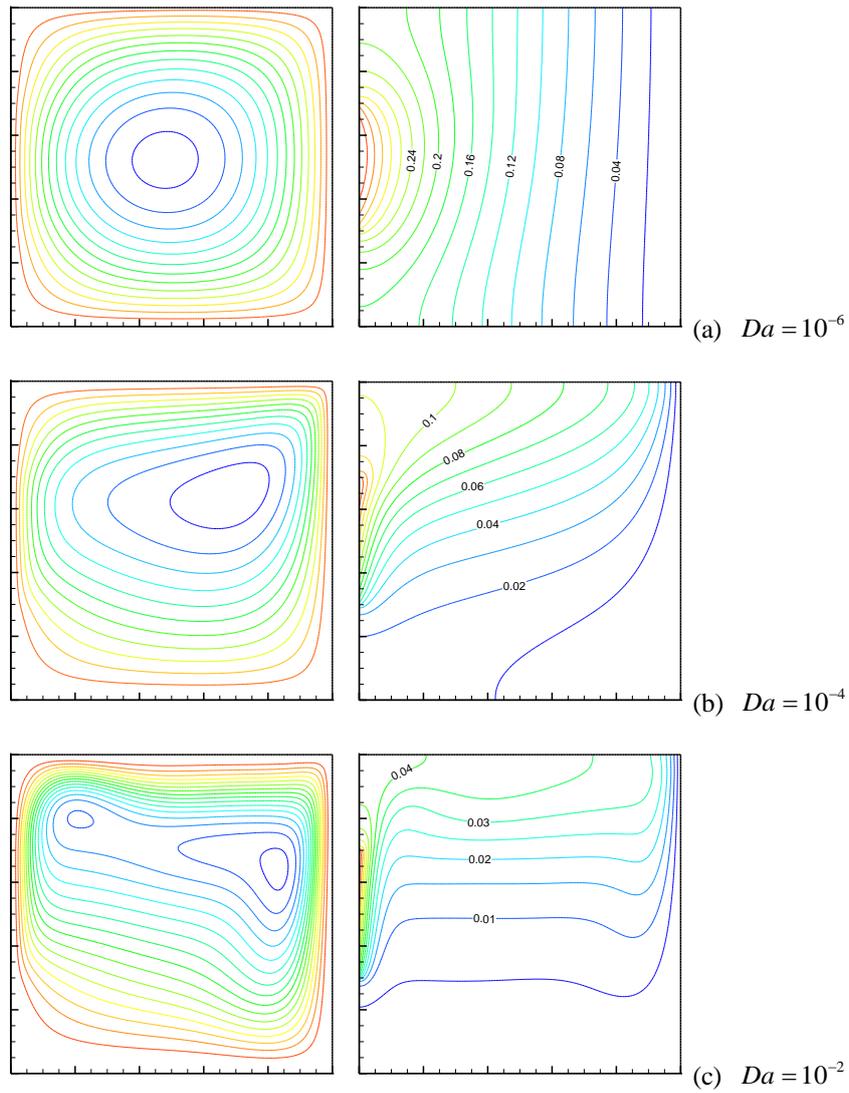

Fig. 8. Streamlines (left) and isotherms (right) for different Darcy numbers: (a) $Da = 10^{-6}$; (b) $Da = 10^{-4}$; (c) $Da = 10^{-2}$.

**Table Captions**

Table 1. Comparisons of the Nusselt numbers for different Darcy numbers.

Table 2. Comparisons of the average Nusselt numbers in the present study with those reported in previous studies.

Table 3. Comparisons of the average Nusselt numbers and the maximum values of the stream function in the present study with the results in Ref. [5].

Table 1. Comparisons of the Nusselt numbers for different Darcy numbers.

| $Da$ | Ref. [6] | Present |
|---|---|---|
| $10^{-3}$ | 5.428 | 5.467 |
| $3\times10^{-3}$ | 5.188 | 5.226 |
| $10^{-2}$ | 4.791 | 4.770 |
| $3\times10^{-2}$ | 4.353 | 4.390 |
| $10^{-1}$ | 3.963 | 3.977 |
| 0.5 | 3.731 | 3.745 |
| Without porous | 3.658 | 3.674 |

Table 2. Comparisons of the average Nusselt numbers in the present study with those reported in previous studies.

| $Ra$ | Ref. [32] | Ref. [33] | Ref. [49] | Ref. [50] | Present |
|---|---|---|---|---|---|
| $10^3$ | - | 1.692 | - | - | 1.710 |
| $10^4$ | 3.216 | 3.215 | 3.037 | 3.163 | 3.211 |
| $10^5$ | 5.782 | 5.787 | 5.760 | 5.882 | 5.768 |

Table 3. Comparisons of the average Nusselt numbers and the maximum values of the stream function in the present study with the results in Ref. [5].

| $Da$ | $\overline{Nu}$ | | $|\psi|_{max}$ | |
|---|---|---|---|---|
| | Ref. [5] | Present | Ref. [5] | Present |
| $10^{-6}$ | 2.97 | 2.92 | 0.28 | 0.27 |
| $10^{-4}$ | 9.28 | 9.21 | 5.44 | 5.69 |
| $10^{-2}$ | 17.51 | 17.09 | 16.09 | 16.38 |